\documentclass[pdflatex,sn-mathphys-num]{sn-jnl}
\usepackage{amssymb}
\usepackage{bm}
\usepackage{amsmath}
\usepackage{placeins}
\usepackage{titlesec}
\usepackage[T1]{fontenc}
\usepackage{textcomp}
\usepackage{etoolbox}
\usepackage{etextools}
\usepackage{comment}
\usepackage{booktabs}
\usepackage{multirow}
\usepackage[justification=raggedright,singlelinecheck=false]{caption}
\usepackage[acronym]{glossaries}
\glsdisablehyper
\makeglossaries
\usepackage[utf8]{inputenc}
\usepackage{hyperref}
\usepackage{graphicx}
\usepackage{dcolumn}
\usepackage{color}
\def  \bea  {\begin{eqnarray}}
\def  \eea  {\end{eqnarray}}

\newcommand{\stkout}[1]{\ifmmode\text{\sout{\ensuremath{#1}}}\else\sout{#1}\fi}

\renewcommand{\thesection}{\Roman{section}}
\renewcommand{\thesubsection}{\Roman{section}.\Roman{subsection}}
\usepackage{setspace}

\usepackage{indentfirst}
\usepackage{amsfonts}
\usepackage{float}
\usepackage{caption}
\usepackage[labelfont=bf]{caption}
\usepackage{subcaption}
\usepackage{geometry}
\usepackage{mathtools}
\usepackage{enumerate}
\usepackage{booktabs}
\usepackage{physics}
\usepackage{multirow}
\usepackage{tabularx}

\usepackage{mathrsfs}
\usepackage{placeins}
\usepackage{grffile}
\usepackage[normalem]{ulem}
\usepackage{tikz} 
\usepackage{mathrsfs}
\usepackage[normalem]{ulem}

\begin{document}

\title[Article Title]{Modelling Dissipative Dynamics of r-mode Instability in Hybrid Stars}

\author[1]{\fnm{Khushbu} \sur{Zala}}\email{khushbu.zalaa001@nmims.in}

\author[1]{\fnm{Sreemoyee} \sur{Sarkar}}\email{sreemoyee.sarkar@nmims.edu}

\affil[1]{\orgname{Mukesh Patel School of Technology Management and Engineering, SVKM's NMIMS University}, 
\orgaddress{\street{Vile Parle (W)}, \city{Mumbai}, \postcode{400056}, \state{Maharashtra}, \country{India}}}


\abstract{Compact star cores reach extreme densities and may contain exotic dense-matter phases. Information about the exotic interiors of rapidly rotating pulsars can be inferred from r-mode oscillations, whose stability is governed by viscous dissipation.  In this work, we model a compact star containing a possible mixed phase of hadronic and quark matter and employ a hybrid statistical framework based on Bayesian inference to infer the dissipation time scales associated with the hybrid phase. Using low-mass X-ray binaries (LMXB) timing observations together with mass–radius constraints from the Neutron Star Interior Composition Explorer (NICER) mission, we estimate the shear and bulk viscosity contributions to r-mode damping for a hybrid star of two layers. Our inference yields shear and bulk viscous dissipation time scales of $\tau_s=(4.99^{+0.49}_{-0.52}) \times 10^8 T^{\frac{5}{3}}$s and $\tau_B= (2.150^{+1.23}_{-0.60}) \times 10^{19} (T^4 10^{-12}+T^2 10^{-6})^{-1}\Omega^{-2}$s respectively. The timescales thus obtained can be implemented to obtain the minima of the star's rotation frequency at 
$\Omega=451.87$ Hz at temperature $T=0.259$ MeV for a hybrid star of mass $1.5$ $M_{\odot}$ and $\Omega=517.47$ Hz at $T=0.234$ MeV for $M=1.75 M_{\odot}$.  We find that the instability window obtained through the inference framework effectively explains the observed stability of millisecond pulsars in both the radio and LMXB populations, particularly for XTE J0929-314 and XTE J1807-294, J0437-4715,  J2124-3358, respectively. These results demonstrate that Bayesian inference combined with r-mode phenomenology provides a powerful and observationally consistent framework for constraining the transport properties of dense hybrid matter.}




\maketitle

\section{Introduction}\label{sec1}

Neutron stars act as natural astrophysical laboratories where dense nuclear matter phases arise under conditions of extreme density and finite temperature. While binary neutron star mergers (BNSMs) and Relativistic Heavy Ion Collisions (RHIC) probe hot and dense matter relevant to the Quantum Chromodynamics (QCD) phase diagram, isolated neutron stars provide insights into cold, dense matter. As we probe deeper inside a neutron star,  at sufficiently high densities and temperatures matter is expected to undergo a deconfinement transition, transforming  hadronic matter to degenerate quark matter. This raises the possibility that compact stellar objects may contain a mixed phase, where quark and hadronic matter coexist. 

A major challenge in determining the internal composition of compact stars from observations stems from significant theoretical uncertainties in the equation of state (EoS) and transport properties of dense matter. For hadronic matter, experimental constraints are restricted to relatively low densities, making extrapolations to neutron star interiors uncertain. The situation is even more challenging for possible quark matter phases, as QCD cannot yet be solved reliably in the non-perturbative regime. In this context, pulsar rotational frequencies and their time derivatives, being the most precisely measured observables, provide a powerful probe of neutron star interiors alongside NICER mass-radius measurements. Extracting meaningful constraints, however, requires identifying robust signatures that connect  macroscopic observables to microscopic physics. R-mode oscillations offer such a possibility. The majority of existing investigations analyze r-modes within the framework of ideal fluid hydrodynamics. However, in regions of significant viscous dissipation, the r-mode oscillations deviate substantially from their ideal, inviscid profiles, introducing enhanced damping mechanisms that effectively truncate r-mode amplitudes. Deep within the neutron star core, where hadronic matter transitions to quark matter through a mixed phase, this r-mode damping arises from the complex interplay between hadronic-quark bulk ($\zeta$) and shear ($\eta$) viscosities. Consequently, mapping the resulting r-mode instability window by calculating the viscous dissipation timescales ($\tau_S, \tau_B$) via these key transport coefficients provides a valuable avenue for identifying the presence of mixed phase in neutron star interiors. However, quantifying the $\zeta$ and $\eta$  within the mixed phase, and subsequently identifying their imprints on the r-mode instability window, still remains  challenging in the realm of nuclear astrophysics.

Several theoretical studies have investigated the microphysical origins of viscosity in neutron star matter, considering both hadronic and quark phases. However, large uncertainties remain in the density regime where a  hadronic matter and quark matter may exist together. Over the last decade a series of works have been done in this direction, authors of Refs.\cite{Madsen:1992sx, Jones:2001ya, Alford:2010gw, Yakovlev:2018jia, PhysRevD.39.3804, Heiselberg:1993cr} have investigated the microphysical and hydrodynamical aspects of $\zeta$  in neutron stars.  In particular, Refs.\cite{Jones:2001ya, Yakovlev:2018jia, Alford:2010gw} examined $\zeta$ in the hadronic phase and its role in dissipating energy within the stellar fluid, while works such as \cite{Madsen:1992sx, Alford:2010gw} focused on the calculation of the $\zeta$ in the quark matter. Recently,  the authors \cite{Alford:2017rxf,Alford:2023gxq,Alford:2021lpp,Alford:2019kdw,Most:2021zvc,Alford:2021ogv,Sedrakian:2022kgj, Sarkar:2024frz} have examined different microphysical characteristics of  $\zeta$ in BNS mergers. 

Detailed studies of $\eta$ in neutron star cores, including Landau damping and in-medium nucleon mass modifications, are presented in Ref.~\cite{Shternin:2008es}. $\eta$ in superfluid phases, accounting for phonon interactions, is explored in Ref.~\cite{Manuel:2012rd}. Transport and relaxation properties of degenerate quark matter are discussed in Ref.~\cite{Heiselberg:1994ms}. $\eta$ in crystalline color-superconducting quark matter is studied in Ref.~\cite{Alford:2014doa}, while transport coefficients in two-flavor superconducting (2SC) quark matter are investigated in Ref.~\cite{Sarkar:2016gib}. The role of shear viscous dissipation in neutron-star mergers is examined in Ref.~\cite{Alford:2017rxf}.  Early estimates of both $\zeta$ and $\eta$ in the hadron-quark mixed phase were presented in Refs.~\cite{Chatterjee:2007qs, Drago:2003wg,Pan:2006dp, PhysRevD.85.103001}, where  unified EoS incorporating both hadronic and quark degrees of freedom was employed.

Building on these studies of viscous properties in dense matter, a number of works have focused on the impact of such dissipation mechanisms on the r-mode instability in compact stars. In Ref.~\cite{Alford:2010fd} r-mode instability has been examined for a wide range of neutron stars, strange stars and hybrid stars assuming a sharp phase transition between hadronic and quark matter. The viscous damping of r-mode oscillations in compact stars containing quark matter is investigated in \cite{Jaikumar:2008kh}, emphasizing the role of $\zeta$ in the instability. Gravitational wave emission from oscillating millisecond pulsars is studied in \cite{Alford:2014pxa}, showing that r-mode oscillations can drive spin-down and constrain the internal composition of compact stars. Observations of low-mass X-ray binaries are used in Ref. \cite{Mahmoodifar:2013quw} to constrain r-mode amplitudes and the associated gravitational wave emission. Recent studies on neutron star r-mode instabilities also used X-ray and UV observations to constrain the thermal and rotational properties of rapidly rotating neutron stars \cite{10.1111/j.1365-2966.2012.21171.x}, while further analyses derived strong upper limits on r-mode amplitudes in millisecond pulsars using X-ray observations \cite{10.1093/mnras/stw3201}. A review of gravitational wave-driven r-mode instability in rapidly rotating neutron stars is presented in \cite{Haskell:2015iia}, discussing observational constraints and additional damping mechanisms.

Motivated by these developments, we  determine the r-mode instability window of a two-layer compact star model with hadron-quark mixed phase deep inside the core of the compact object. This problem is inherently ill-posed due to the poorly constrained nature of the high-density EoS, along with significant uncertainties in both observational data and theoretical inputs. In this context, Bayesian inference provides a natural and rigorous framework to address these challenges by incorporating prior physical constraints—such as thermodynamic stability together with  X-ray measurements, and pulsar timing. By incorporating observational constraints on neutron star mass-radius with pulsar timing data, we infer the r-mode instability window by calculating dissipation timescales associated with $\eta$ and $\zeta$. For this  we consider a two-layer compact object consisting of baryonic matter in the outer layer ($n_B< 2.23 n_0$) ($n_B$ baryonic density, $n_0$ nuclear saturation density) and an inner mixed phase where hadronic and quark matter coexist for the innermost regime of baryonic density higher than the above mentioned transition density. For the hadronic phase, we employ relativistic mean-field DDME2 EoS \cite{Sarkar:2024frz, Lalazissis:2005de}, while the mixed-phase EoS at intermediate to high densities is constructed within the Bayesian framework. Apart from high density EoS, we characterize the dissipative time scales through the inferred dissipative coefficients namely $\tilde S, \tilde V  $ and $\tilde W$ which we define later in the paper. These time scales are subsequently employed to construct the r-mode instability window. Our analysis provides a self-consistent description of the transition in $\tau_S$ and $\tau_B$   from the hadronic phase to the mixed phase, thereby offering a potential observational signature of phase transitions in hybrid star interiors through rotational instabilities.

The paper is organized as follows in Section: II we present the formalism of r-mode oscillation with the help of different viscous time scales.  In Section III, we present the methodology of formulating hybrid star  $r-mode$ instability curve and in the Section IV results of r-mode oscillation in hybrid star have been presented. Finally, in Section V, we  summarize and conclude.

\section {Dissipation time scales and r-mode oscillation}
\label{sec2}

Fluid in a rotating neutron star becomes unstable to a mode that couples to gravity when viscous damping is absent. These non-radial oscillation modes emit gravitational radiation, carrying away the star’s angular momentum. As the modes grow, the neutron star is expected to spin down rapidly a phenomenon known as the r-mode instability. However, this growth is constrained by viscous dissipation, primarily through bulk and shear viscosity. Both the coefficients   together determine whether r-modes are suppressed or amplified.

The amplitude of the r-mode oscillations varies over time according to the expression \( e^{i\omega t - t/\tau} \), where \( \omega \) denotes the real component of the r-mode frequency, and \( 1/\tau \) represents its imaginary component.  The quantity \( 1/\tau \) can be expressed as \cite{Alford:2010fd}:
\begin{align}
\frac{1}{\tau(\Omega, T)} = \frac{1}{\tau_G(\Omega)} + \frac{1}{\tau_B(\Omega, T)} + \frac{1}{\tau_S(T)}
\label{eq:tau_decomposition}
\end{align}
The minimum position of the r-mode instability curve corresponds to the temperature ($T_{min}$) at which the angular velocity of the stellar object reaches its lowest value, arising from the competition between gravitational radiation, which drives the instability, and viscous dissipation, which suppresses it. At low and high temperatures, viscous effects dominate and stabilize the mode, while at intermediate temperatures damping is least effective, making the star most susceptible to instability. This minimum is determined by the condition $d\Omega_c/dT=0$
 and identifies the most unstable region of the star, providing a sensitive probe of the underlying microphysics, including viscous dissipation and the equation of state. If the star consists hybrid matter where, each component is characterized by its own viscosity, the total viscous damping can be obtained  from the combined contributions of all components.  The timescale for r-mode amplification driven by gravitational wave emission is given below \cite{Lindblom1998},
\begin{align}
\frac{1}{\tau_G} = - \frac{32\pi 2^{4}}{15^2}
\left( \frac{4}{3} \right)^{6} 
\tilde{J}_2 G M R^{4} \Omega^{6}.
\label{eq:tau_g}
\end{align}
{In the above equation $\tilde{J_2} \equiv \frac{1}{M R^{4}} \int_0^R \rho(r)\, r^{6} \, d r$ is radial integral constant. Here, $\rho(r)$ represents the star’s radial energy density profile obtained from the solution of the TOV equations, $G$ is the gravitational constant, and $M$, $R$ denote the stellar mass and radius, respectively.

 The damping time scale of r-mode due to shear viscosity is given by $\tau_S^{-1}=(5/\tilde{J}_2 M R^{4})\int_0^R  \eta r^{4} d r$. Using the parametrization $\tilde{\eta}=\eta T^{-5/3}$, this relation can be rewritten as \cite{Alford:2010fd}, 
\begin{equation}
\frac{1}{\tau_S} = \frac{5\tilde{S}_2\Lambda^{\frac{14}{3}}_{QCD}R}{\tilde{J}_2MT^{\frac{5}{3}}}.
\label{eq:tau_s}
\end{equation}

In this expression, the dimensionless coefficient $\tilde{S}_2$, which encodes the dependence on the medium’s shear viscosity, is defined as 
$\tilde{S}_2 \equiv 1/(R^{5} \Lambda_{Q C D}^{\frac{14}{3}} )\int_{0}^{R} \tilde{\eta} r^{2 m} d r$,
where $\Lambda_{QCD}$ denotes the characteristic QCD scale, taken to be $1$ GeV.

The damping timescale of the r-mode caused by $\zeta$ is described by the \( \tau_B \), which accounts for the time scale related to the energy dissipation due to the periodic compression and expansion of the stellar fluid. The inverse of the bulk viscous timescale is expressed as
$\tau_{B}^{-1} = (4\int d^{3}x | \Delta n_B/\bar{n_B}|^{2} \zeta)/(9\alpha^{2} \tilde{J}_{2} M R^{2}) $, where \( \alpha \) is the dimensionless amplitude of the r-mode,  \( \zeta \) denotes the bulk viscosity, $\Delta n_B$ is the fluctuation in the baryon density \cite{Alford:2010fd}. In the limit when the non-linear component of the periodic fluctuation in chemical
potential can be neglected, the inverse of this timescale is expressed as,

\begin{align} \frac{1}{\tau_{B}} = \frac{96}{3402} \frac{R^5 \Omega^3}{\tilde{J}_2 M} \, {\cal D} \left( \frac{3T^{\delta}}{2 \Omega} \right) \label{equation} \end{align}

where, the function \({\cal D}{}\) captures the star's microphysical characteristics, including the medium's reaction to density perturbations.  $\delta$ is an important exponent that encodes the  information about various types of weak interactions. In particular, $\delta = 4$ corresponds to the hadronic direct Urca process (DURCA) which we consider for the baryonic medium. Now, writing $\mathcal{D}$ explicitly we obtain, 
\begin{align}
\mathcal{D}\left(\frac{3T^{4}}{2\Omega}\right) \equiv \frac{3T^4}{2\Omega R^3} \int_{R_i}^{R_o} dr\, r^2 \, \frac{A_{h, m}^2 C_{h, m}^2 {\tilde\Gamma_{h,m}}}{1 + \tilde\Gamma_{h,m}^2 B_{h, m}^2 b^2} \left( \frac{r}{R}\right)^9,
\end{align}

where, \( R_i \), \( R_o \) are the radial extent of the region contributing to the damping, typically corresponding to distinct internal layers like the core or crust. ${A_{h,m}}$, ${B_{h,m}}$, and ${C_{h,m}}$ represent thermodynamic susceptibilities of hadronic matter (h) or in the mixed phase (m). The coefficient ${A_{h,m}}$ is related to the inverse square of the sound speed and is given by
${A_{h,m}=\left(\partial \rho/\partial P\right)}|_{h,m}$ ($P$ denotes pressure). ${\tilde{\Gamma}_{h,m}}$ denotes the dimensionless weak interaction rate in the hadronic or mixed-phase regimes. The susceptibilities characterize the response of the chemical potential imbalance to perturbations in the number density and the composition. The thermodynamic susceptibilities corresponding to different phases of dense matter are given as follows \cite{Alford:2010gw}:
\bea
C_{h,m} = \bar{n}_B 
\left.\frac{\partial\mu_{\Delta(h,m)}}{\partial n_B}\right|_{x}, \hskip 12pt  B_{h,m} = \frac{1}{\bar{n}_B} \left.\frac{\partial\mu_{\Delta(h,m)}}{\partial x}\right|_{\bar{n}_B}.
\label{eq:ch_bh}
\eea

In the above expressions, $\mu_{\Delta}$ denotes chemical potential imbalance due to deviation from beta equilibrium. The relaxation of ${\mu_\Delta}$ back to zero determines the rate at which the chemical composition of matter readjusts to pressure perturbations. 

In the asymptotic limit of the stellar oscillation frequency ($\omega$), one can derive analytic formulae for $\tau_B$. In the high-frequency domain, where $f = B\tilde{\Gamma} T^{\delta}/\omega \ll 1$, as well as in the opposite limit, the corresponding characteristic dissipation time scales are given by \cite{Alford:2010fd},
\begin{align}
\frac{1}{\tau_{B}^{<}} & \xrightarrow[f\ll1]{}\frac{32}{1701\kappa^{2}}\frac{\Lambda_{QCD}^{9-\delta}\tilde{V}_{2}R^{5}\Omega^{2}T^{\delta}}{\Lambda_{EW}^{4}\tilde{J}_{2}M}, \hskip 6pt \frac{1}{\tau_{B}^{<}} \xrightarrow[f\gg1]{}\frac{32}{1701}\frac{\Lambda_{EW}^{4}\Lambda_{QCD}^{\delta-1}\tilde{W}_{2}R^{5}\Omega^{4}}{\tilde{J}_{2}MT^{\delta}}\label{eq:bulk-damping-time-high}.\end{align}
  In the above equation if $f \ll 1$, the equilibration rate is suppressed, whereas in the complementary regime the equilibration rate increases. 
  Since bulk viscosity arises from weak interactions, weak interaction normalization scale is introduced and typically taken to be 
$\Lambda_{EW}=100GeV$. These normalization scales are employed only to render the constants dimensionless. Finally, it is important to note that, in the expressions of $\tau_B$  the underlying microscopic dissipation mechanism is encoded in a set of constants $\tilde V, \tilde W$. These two constants primarily depend upon the $\zeta$ of the medium.

For a compact star composed of two distinct layers, each region consists of different phase of dense matter, characterized by its own viscous properties. Accordingly, the total viscous damping rate can be written as the sum of the contributions from the individual layers, 
\[
\tau_{v}^{-1} = \sum_{i} \tau_{i,v}^{-1}, \quad i = h, m,\quad v = S, B,
\]
where \(v = S, B\) correspond to the shear and bulk viscous contributions. In the next section we present methodology to obtain $\tau_S$ and $\tau_B$ in different phases.

\section{Methodology}\label{sec3}

The study of oscillation modes in compact stars and their associated dissipation mechanisms requires an accurate determination of the stellar equilibrium configuration. In this section, we describe the stellar models employed in the subsequent analysis. We consider a hybrid star characterized by a two-layer internal structure, allowing for the existence of a hadron–quark mixed phase in the stellar core. The low-density outer layer, corresponding to baryon number densities $n_B<2.23n_0$, is composed of  hadronic matter consisting of neutrons, protons, and electrons under degenerate condition ($T\ll \mu_i$) ($\mu_i$ is the chemical potential of each constituent of the plasma). The hadronic phase is described within the framework of the relativistic mean-field (RMF) theory using the DDME2 EoS \cite{Lalazissis:2005de}. For densities exceeding the transition threshold, the hadronic matter undergoes a phase transition to a hadron–quark mixed phase. This mixed phase extends to higher core densities, within which hadronic and quark degrees of freedom coexist in thermodynamic equilibrium.

Since the EoS of dense matter at supranuclear densities remains poorly constrained, we employ a Bayesian inference framework to model the mixed phase, following Ref. \cite{Greif:2018njt}. Once the EoS is constructed, we evaluate the relevant dissipative time scales associated with the mixed phase, namely the shear-viscous and bulk-viscous damping time scales, $\tau_{m,S}$ and $\tau_{m,B}$, respectively. In particular, we infer the effective parameters governing these dissipation mechanisms, such as  $\tilde S$, $\tilde V$, $\tilde W$, using Bayesian inference techniques \cite{10.1093/mnras/stw3201, 10.1111/j.1365-2966.2012.21171.x}. The resulting dissipation time scales are subsequently employed to determine the r-mode instability window of hybrid stars, and the theoretical predictions are compared with available observational constraints. The basic formalism of the statistical inference framework employed in this work is presented below.

We employ a Bayesian inference framework to constrain the model parameters, enabling a systematic incorporation of both theoretical inputs and observational data. The prior information on the parameters is updated by a likelihood function, which quantifies an understanding between model predictions and observations. This framework allows a consistent estimation of parameter uncertainties and correlations. For a given set of observed data \(D\), Bayes’ theorem can be expressed as\cite{speagle2019},
\begin{equation}
 P(\theta|D) = {\frac {P(D| \theta) P(\theta)}{P(D)}}, \label{Bayes} 
\end{equation}
where, $P(\theta|D)$ is the posterior distribution, $P(D| \theta)$ is the likelihood function and $P(\theta)$ is the prior distribution function. In the current paper we employ the Bayesian inference mechanism  based upon the UltraNest nested sampling algorithm which provides a robust computational framework for estimating the Bayesian evidence for different EoS models as well as transport model \cite{Huang:2024rfg}.

\subsection*{Modeling the EoS of the Hybrid Star}

 The equilibrium structure of a compact star is obtained by solving the general relativistic Tolman–Oppenheimer–Volkoff equations with the EoS as input. For this, we construct a unified EoS comprising the relativistic mean-field model DDME2 for hadronic matter below transition density ($n_B< 2.23n_0$), while the higher-density region is described by  sound speed parametrization technique \cite{Greif:2018njt},
\begin{equation}
     c^2_s(x)/c^2=a_1e^{-\frac{1}{2}(x-a_2)^2/a_3^2}+a_6+\frac{\frac{1}{3}-a_6}{1+e^{-a_5(x-a_4)}}
     \label{sound_speed}
\end{equation}
where $x=\varepsilon/(m_Nn_0)$ and the nucleon mass. In this formulation, the parameters $a_1 \cdots a_6$ are treated as free variables and are allowed to vary over a broad domain. This parametrization  permits the generation of a wide spectrum of EoSs, ranging from very soft to very stiff. For two different central densities as mentioned below in the table we obtain two different hybrid stars. The corresponding values of $M$, $R$ and the Kepler Frequency ($\Omega_K=\sqrt{GM/R^3}$) are listed in the Table~(\ref{table1}) below.
\begin{table}[h]
\centering
\renewcommand{\arraystretch}{1.4}
\begin{tabular}{lcccccc}
\hline
  &  \textbf{M ($M_\odot$)} & \textbf{R (km)} & \textbf{Central density} & \textbf{$\Omega_k$($s^{-1}$)}\\ 
\hline

Hybrid Star & $1.5^{+0.075}_{-0.079}$ & 11.00 & 4.89$n_0$ & $1.22\times10^{4}$\\ \hline

Hybrid Star&  $1.75^{+0.121}_{-0.172}$ & 10.3 & 6.66$n_0$ & $1.45\times10^{4}$\\ \hline

\end{tabular}
\caption{Results of the hybrid star configuration}
\label{table1}
\end{table}
With this hybrid star configuration in hand we describe the modeling of dissipation constants $\tilde S_2, \tilde V_2, \tilde W_2$ and consequently $\tau_S$ and $\tau_B$ in the next section.

\subsection*{Dissipation constants and related time scales}

To determine the r-mode instability window of hybrid stars, it is necessary to evaluate the dissipative timescales associated with the $\eta$ and $\zeta$, respectively. In the present study, we consider a two-layer model of baryonic matter that undergoes a transition to a quark–hadron mixed phase. The dissipation constants, and thus the characteristic timescales $\tau_S$ and $\tau_B$, are obtained by evaluating separately the contributions from the purely baryonic phase and from the mixed-phase region.  We begin by evaluating the time scales  in hadronic matter.

\subsubsection*{Dense baryonic matter}
In the outermost layer of hadronic matter, located just below the transition density, the composition is dominated by neutrons, protons and electrons. In the supranuclear density regime, the electronic contribution to $\eta$ ($\eta_e$) dominates   over the hadronic contribution; therefore, only the electronic component is included in the present analysis \cite{Alford:2010fd}. 
The $\tau_{h,S}$ is calculated using the electron shear viscosity coefficient given by $\eta_{e}=4.26\times10^{-26}(x\, n_B)^{14/9}T^{-5/3}\mathrm{g}\mathrm{cm^{-1} s^{-1}}$, where $x$ denotes the proton fraction and $n_B$ represents the baryon number density. Employing $\eta_e$, we obtain $\tilde S_2$  and hence $\tau_S$ following Eq.(\ref{eq:tau_s}).

In the bulk viscous channel the $\tau_B$, is evaluated using the thermodynamic susceptibilities  obtained from the DDME2 equation of state \cite{Alford:2010fd, Sarkar:2024frz} and also the rate  for direct Urca process as the dominant weak interaction channel as given in the Eq.(\ref{eq:bulk-damping-time-high}). The corresponding direct Urca interaction rate is given by $\Gamma_h=5.24\!\times\!10^{-15}\left(x n_B/n_{0}\right)^{\!\frac{1}{3}} T^4$ \cite{Alford:2010fd}. Using the resulting bulk viscosity coefficient $\zeta_h$, we compute the dimensionless coefficients $\tilde{V}$ and $\tilde W$, which are then used to evaluate the bulk viscosity damping time scale $\tau_{h,B}$.  We now present the methodology  to estimate $\tau_{S,m}$ and $\tau_{B,m}$ within  Bayesian inference framework.

\subsubsection*{Mixed Phase}

In the present section the  $\tau_{S,m}$ and $\tau_{B,m}$ are evaluated within  a hybrid methodology that combines statistical inference techniques with analytical formulations. In particular, the dimensionless coefficients $\tilde J_2$, $\tilde S_2$, $\tilde V_2$ and $\tilde W_2$  of the mixed phase, are constrained using Bayesian inference  with the help of the available observational data \cite{10.1093/mnras/stw3201, 10.1111/j.1365-2966.2012.21171.x}. These inferred coefficients effectively contain information regarding the  $\eta$ and the $\zeta$ of the mixed phase. 

Once the posterior distributions of the relevant coefficients are obtained, the corresponding dissipation time scales are calculated using Eqs.(\ref{eq:tau_s}) and (\ref{eq:bulk-damping-time-high}). These time scales are subsequently employed to determine the r-mode instability window of hybrid stars.

To obtain the viscous parameter estimation we first consider  uniform  priors for the r-mode viscous parameters $\tilde{S}$, $\tilde{J}$,$\tilde{V}$ and $\tilde{W}$. Specifically, we consider the prior distributions  as $\tilde{S} \sim \mathcal{U}[2.19 \times 10^{-6},\, 5.24 \times 10^{-6}]$,$\tilde{J} \sim \mathcal{U}[1.93 \times 10^{-2},\, 2.00 \times 10^{-2}]$, $\tilde{V} \sim \mathcal{U}[1.38 \times 10^{-10},\, 1.07 \times 10^{-3}]$, $\tilde{W} \sim \mathcal{U}[1.32 \times 10^{-6},\, 1.03 \times 10^{-2}]$. The reason for the choice of the prior estimation lies from the existing data of pure neutron star, strange star or hybrid star \cite{Alford:2010fd}. 

The likelihood function is constructed as the sum of independent Gaussian contributions from the mass-radius constraint \cite{Greif:2018njt} and the spin frequency ($\nu$)-temperature observations of neutron stars \cite{10.1093/mnras/stw3201, 10.1111/j.1365-2966.2012.21171.x}. For a given 
the mass-radius likelihood is  given by \cite{Greif:2018njt},
\begin{equation}
\mathcal{L}_{\mathrm{MR}} =
 \frac{\sigma_M\sigma_R}{2}exp \left[
\frac{(M - \mu_m)^2}{\sigma_M^2}
+
\frac{(R - \mu_r)^2}{\sigma_R^2}
\right],
\end{equation}
where $M$ and $R$ represent the model predictions, $\mu_m$ and $\mu_r$ denote the mean observational values, and $\sigma_M$ and $\sigma_R$ correspond to the 5\% uncertainties of the respective distributions in the data \cite{Typel:2013rza}.  In addition, we incorporate observational constraints on the $\nu-T$ of the neutron stars. 
The likelihood based on spin frequency and temperature measurements is given by,
\begin{equation}
\mathcal{L}_{\mathrm{\nu,T}} =
 \frac{\sigma_\nu\sigma_T}{2}exp \left[
\frac{(\nu_{\mathrm{model}} -\mu_{\mathrm{\nu}})^2}{\sigma_\nu^2}
+
\frac{(T_{\mathrm{model}} - \mu_{\mathrm{T}})^2}{\sigma_T^2}
\right],
\end{equation}
where $\mu_{\nu}$ and $\mu_{T}$ denote the mean values of the observed spin frequency and temperature, while $\nu_{\mathrm{model}}$ and $T_{\mathrm{model}}$ are the corresponding model predictions.  $\sigma_\nu$ and $\sigma_T$ represent the associated observational uncertainties obtained from the data presented in \cite{10.1093/mnras/stw3201, 10.1111/j.1365-2966.2012.21171.x}.
The total likelihood is then obtained by multiplying the independent contributions,
\begin{equation}
\mathcal{L}_{\mathrm{tot}} =
\mathcal{L}_{\mathrm{MR}} \times
\mathcal{L}_{\mathrm{\nu,T}}.
\end{equation} 
After defining the prior and likelihood functions, we employ the nested sampling algorithm, \texttt{UltraNest}, to construct the posterior distribution. This procedure enables an efficient exploration of the parameter space by focusing on regions of higher likelihood, and gives an estimate of the Bayesian evidence. From the obtained posterior samples one can then obtain the distributions for the model parameters, including their uncertainties and correlations.

\section{Results}

In this section, we quantify the  $\tau_S$, $\tau_B$, using the inferred dimensionless coefficients and subsequently determine the corresponding r-mode instability curves for hybrid stars. We first present the posterior distributions of the dimensionless r-mode viscous and equilibrium parameters, $\tilde{S}$, $\tilde{J}$, $\tilde{V}$ and $\tilde{W}$, through corner plots shown in Fig.~(\ref{Cornorplot}). These plots illustrate  constraints on the individual parameters and the pairwise correlations among them for hybrid star configurations with masses $1.5 M_{\odot}$ (left panel) and $1.74 M_{\odot}$ (right panel).

\begin{figure}[H]
\centering

\begin{subfigure}{0.5\textwidth}
\centering
\includegraphics[width=\linewidth]{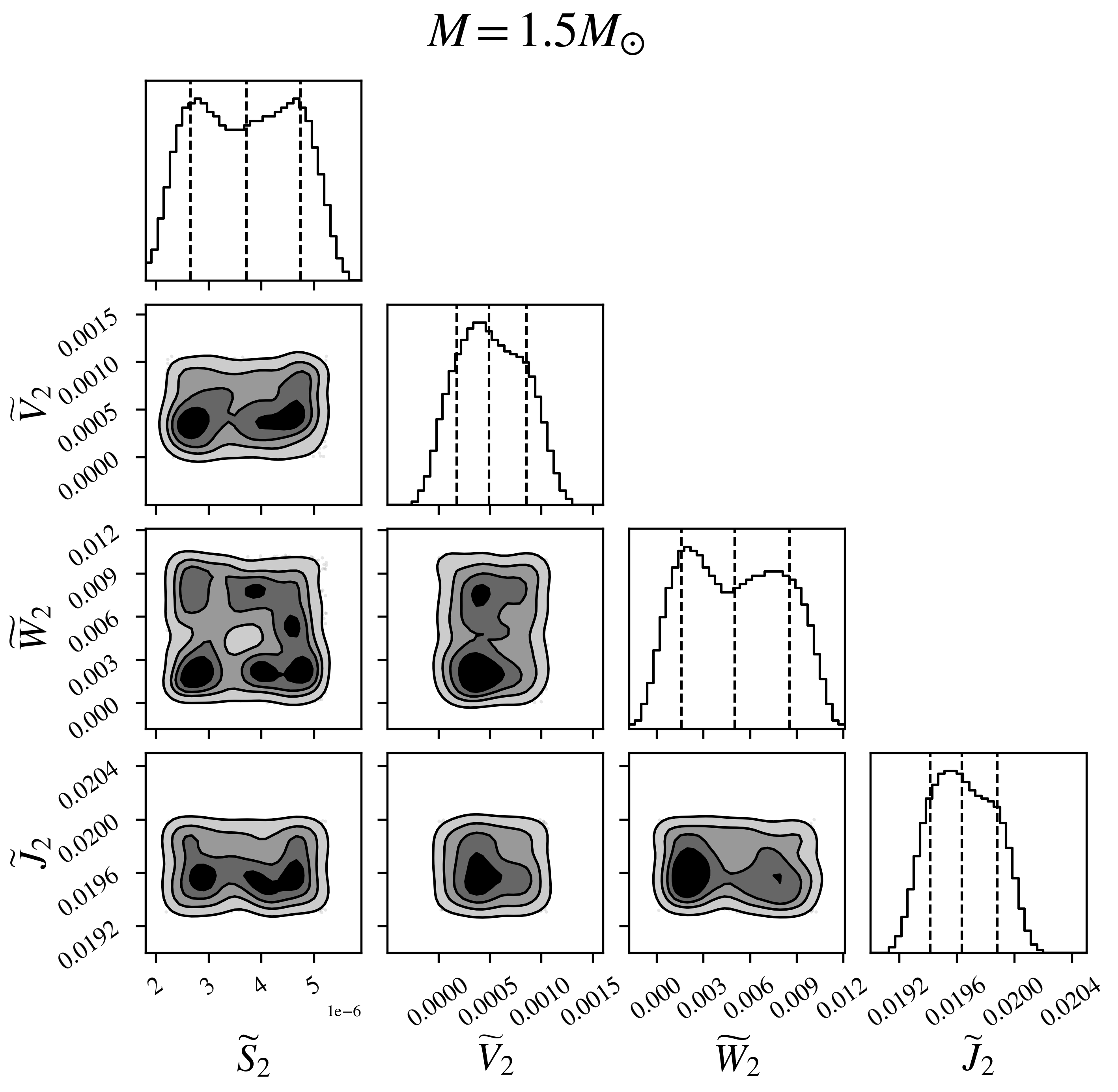}
\caption{}
\end{subfigure}%
\begin{subfigure}{0.5\textwidth}
\centering
\includegraphics[width=\linewidth]{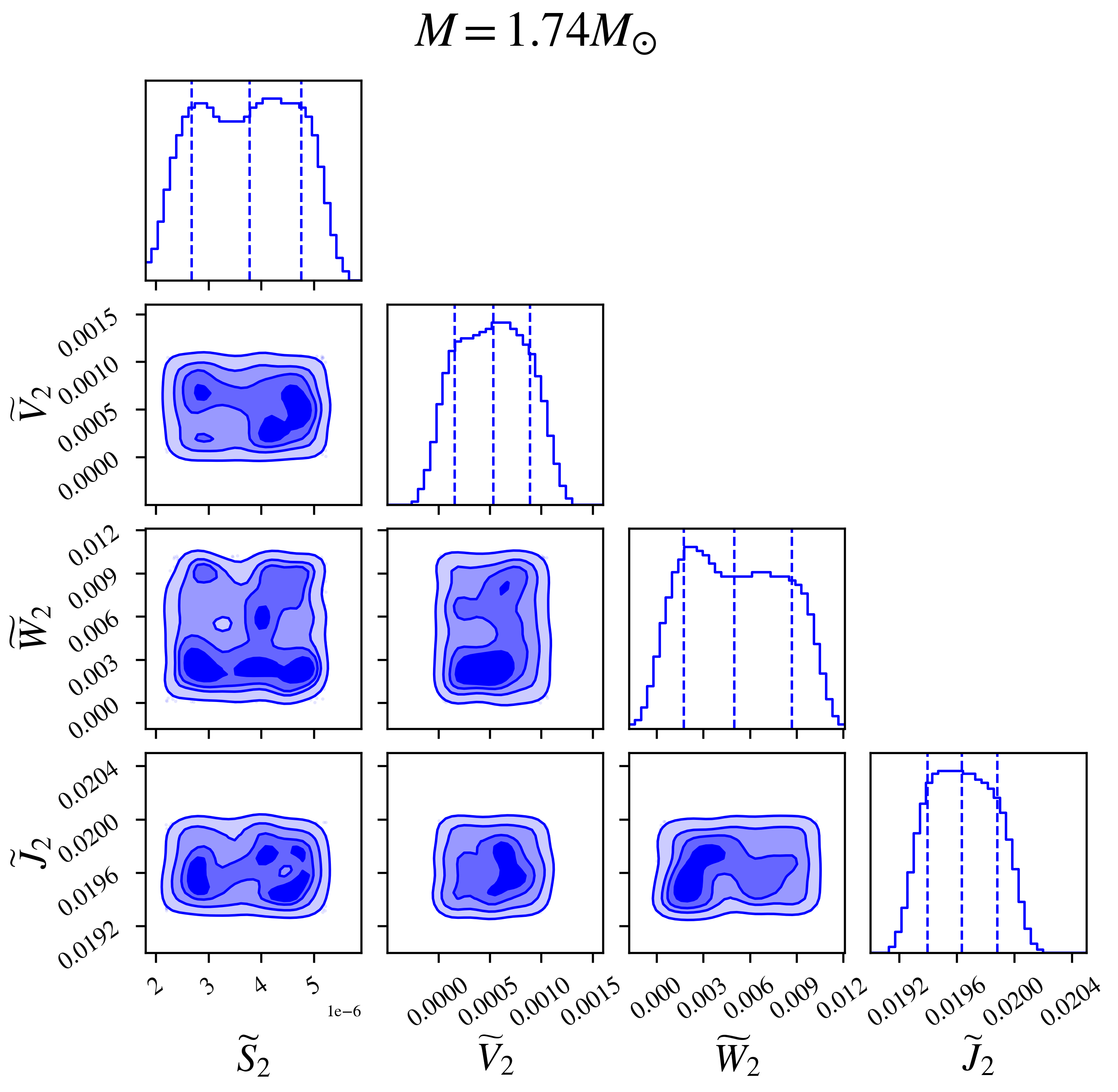}
\caption{}

\end{subfigure}

\caption{Posterior distributions of the r-mode parameters $\tilde{S_2}$,$\tilde{V_2}$,$\tilde{W_2}$ and $\tilde{J_2}$, obtained from the Bayesian inference.}
\label{Cornorplot}
\end{figure}

In the  Fig.(\ref{Cornorplot}), the diagonal panels display the one-dimensional posterior distributions, while the off-diagonal panels show the two-dimensional joint posterior distributions with contours denoting confidence regions. The vertical dashed lines mark the posterior medians and their corresponding 1$\sigma$ credible intervals. Moderate correlations are observed among the  parameters, particularly between $\tilde S_2$, $\tilde V_2$, $\tilde{W_2}$, reflecting the coupled role of $\eta$ and $\zeta$ in determining the instability boundary in the frequency–temperature plane. In contrast, the correlations involving the parameter $\tilde{J_2}$ remain comparatively weak. Compared to the $1.5M_\odot$ configuration, the posterior contours for the $1.74M_\odot$
 star appear narrower and more localized, indicating more stringent constraints on the transport parameters for hybrid star of higher mass. This behavior suggests that the observational frequency–temperature data are sensitive to the dense core dissipation properties in more massive hybrid stars, leading to improved constraints on the underlying transport mechanisms. The mean and   the estimated errors of the constants are presented in the Table (\ref{table2}).
\begin{table}[h]
\centering
\resizebox{\textwidth}{!}{   
\begin{tabular}{lcccc}
\hline
Phase & $\tilde{J}_2$ & $\tilde{S}_2$ & $\tilde{V}_2$ & $\tilde{W}_2$ \\
\hline
Hadron  & $2.51\times10^{-2}$ & $1.4070\times10^{-4}$ & $9.421\times10^{-4}$ & $1.904\times10^{-6}$ \\
\hline
Hybrid  (1.5$M_\odot$)& $(1.965\times 10^{-2} \pm 2.47\times10^{-4})$ & $(3.757 \pm 1.05)\times10^{-6}$ & $(5.3129 \pm 3.57)\times10^{-4}$ & $(4.975 \pm 3.45)\times10^{-3}$ \\
\hline
Hybrid  (1.74$M_\odot$)& $(2.016\times 10^{-2} \pm 2.26 \times10^{-4}$) & $(3.6018 \pm 1.03)\times10^{-6}$ & $(5.266 \pm 3.78)\times10^{-4}$ & $(5.030 \pm 3.62)\times10^{-3}$ \\
\hline
Quark  & $2.92\times10^{-2}$ & $1.099\times10^{-7}$ & $1.67\times10^{-8}$ & $2.8022\times10^{-5}$ \\
\hline
\end{tabular}
} 
\caption{Statistical Inference of $\tilde{J}_2$, $\tilde{S}_2$, $\tilde{V}_2$ and $\tilde{W}_2$}
\label{table2}
\end{table}

\begin{figure}[H]
\centering

\begin{subfigure}{0.5\textwidth}
\centering
\includegraphics[width=\linewidth]{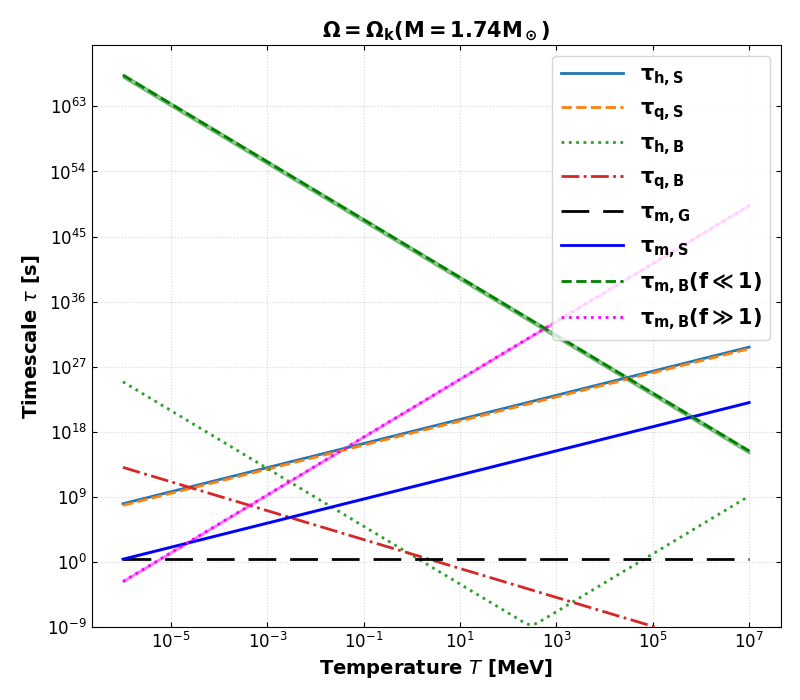}

\end{subfigure}%
\begin{subfigure}{0.5\textwidth}
\centering
\includegraphics[width=\linewidth]{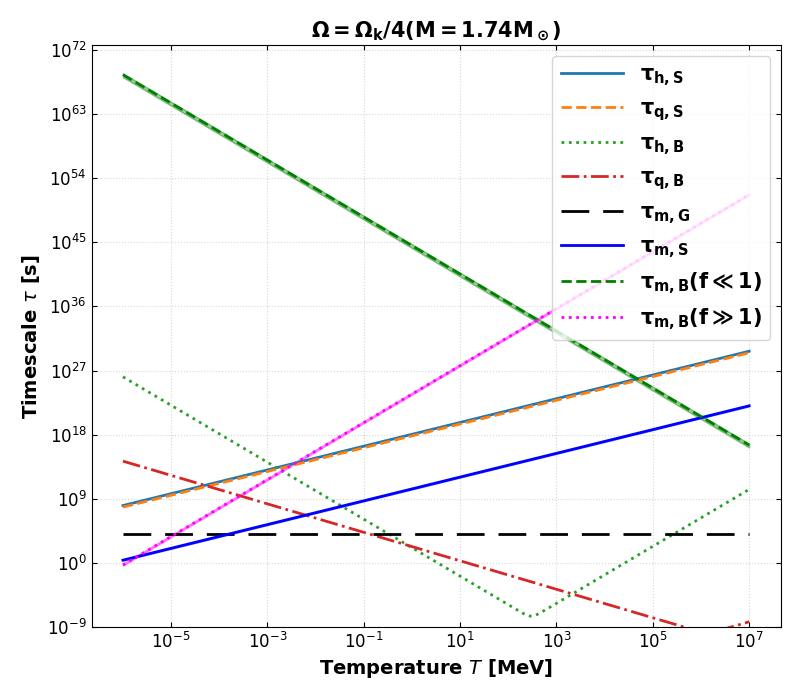}

\end{subfigure}

\caption{Variation of $\tau$ with $T$ }
\label{Timescale}
\end{figure}
\renewcommand{\arraystretch}{1.6}
\setlength{\extrarowheight}{3pt}
\setlength{\tabcolsep}{8pt}

\begin{table}[h]
\centering

\begin{tabular}{|c|c|c|}
\hline
$\tau [s]$ 
& $M = 1.5\,M_\odot,\ R = 11\,\text{km}$ 
& $M = 1.74\,M_\odot,\ R = 13\,\text{km}$ \\
\hline

$\tau_s(T)$ 
& $\left(4.99^{+0.49}_{-0.52}\right)\times 10^{8}\, T^{5/3}$ 
& $\left(4.04^{+0.59}_{-0.6}\right)\times 10^{8}\, T^{5/3}$ \\
\hline

\shortstack{$\tau_B (T,\Omega)$ \\[2pt] $(f \ll 1)$}
& $\left(2.150^{+1.23}_{-0.60}\right)\times 10^{19}
\left[
\frac{1}{T^{4}\,10^{-12} + T^{2}\,10^{-6}}
\right]\Omega^{-2}$
& $\left(3.001^{+1.75}_{-1.66}\right)\times 10^{20}
\left[
\frac{1}{T^{4}\,10^{-12} + T^{2}\,10^{-6}}
\right]\Omega^{-2}$ \\
\hline

\shortstack{$\tau_B (T,\Omega)$ \\[2pt] $(f \gg 1)$}
& $\left(6.38^{+3.84}_{-1.21}\right)\times 10^{9}
\left[
\frac{T^{4}}{10^{12}} +Ss \frac{T^{2}}{10^{6}}
\right]\Omega^{-4}$
& $\left(6.4^{+3.76}_{-1.08}\right)\times 10^{8}
\left[
\frac{T^{4}}{10^{12}} + \frac{T^{2}}{10^{6}}
\right]\Omega^{-4}$ \\
\hline

\end{tabular}
\caption{Dissipation time scales for two different stellar configurations}
\label{timescale}
\end{table}

The Fig.~(\ref{Timescale}) shows the variation of the different dissipation and gravitational radiation timescales with temperature $T$ for a hybrid star of mass $M=1.74M_\odot$. The left panel corresponds to the Kepler angular velocity $\Omega=\Omega_k$, while the right panel corresponds to a slower rotation rate $\Omega=\Omega_k/4$. The solid  and small-dashed monotonic curves represent the  $\tau_{h,S}$ and  $\tau_{q,S}$, respectively. The dotted and dash-dotted non-monotonic curves represent the timescales associated with $\zeta$, $\tau_{h,B}$, and $\tau_{q,B}$, respectively. The long-dashed horizontal curve represents the  gravitational radiation timescale  $\tau_{m,G}$ for the hybrid star. The solid shaded band represents the hybrid shear viscosity timescale $\tau_{m,S}$, the dashed shaded band corresponds to the hybrid bulk viscosity timescale $\tau_{m,B}$  in the regime $(f\ll1)$, and the large-dotted shaded band represents the hybrid bulk viscosity timescale $\tau_{m,B}$  in the regime $(f\gg1)$. The time scales related to the hybrid star are monotonic here, since we have considered the asymptotic domain. Because $\eta$ dominates and stabilizes the star at low temperatures, and $\zeta$ dominates at high temperatures, they create a window at intermediate temperatures where damping is least effective. For purely hadronic matter the $\tau_{h, S}$ and $\tau_{q,S}$ intersection point lies at $T=9.47\times 10^{-4}$ MeV, whereas for the quark matter the minima lies at $T=2.67\times 10^{-5}$ MeV. The inferred minima for the hybrid star is at $T=37.98$ MeV. The summary of the time scales obtained are listed in the Table (\ref{timescale}).

\begin{figure}[H]
\centering

\begin{subfigure}{0.5\textwidth}
\centering
\includegraphics[width=\linewidth]{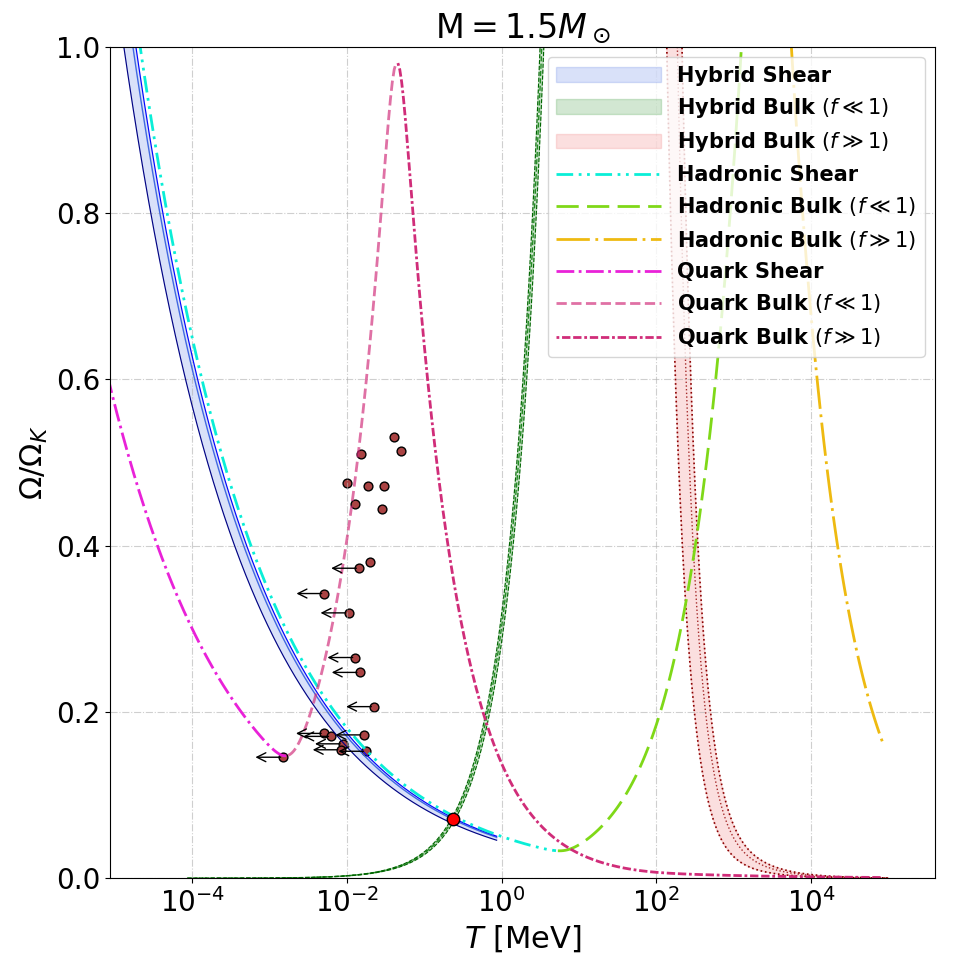}

\end{subfigure}%
\begin{subfigure}{0.5\textwidth}
\centering
\includegraphics[width=\linewidth]{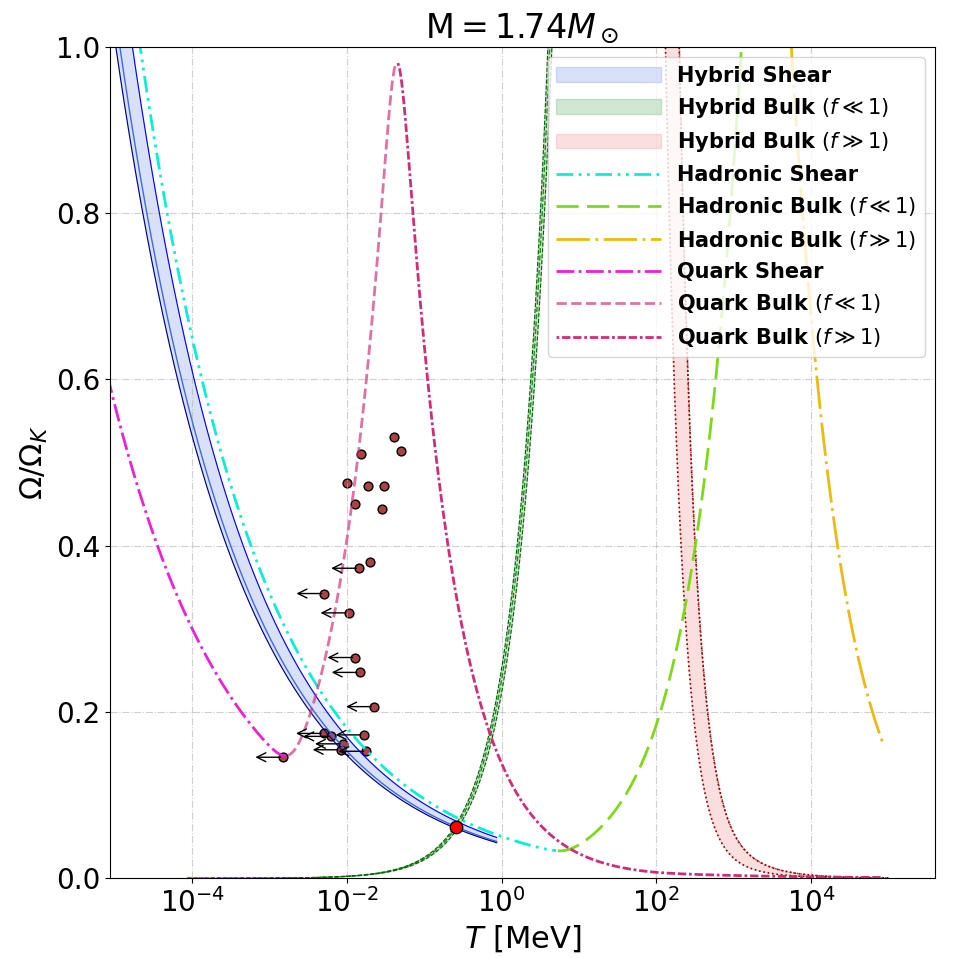}

\end{subfigure}

\caption{Variation of $\Omega/\Omega_K$ with $T$ (MeV)}
\label{rmode}
\end{figure}

In the Fig.(\ref{rmode}) the angular frequency normalized to the Keppler frequency ($\Omega/\Omega_K$) with the mean values and the upper and lower bounds of the  parameters $\tilde{S_2}$,$\tilde{V_2}$,$\tilde{W_2}$ and $\tilde{J_2}$  have been plotted with  $T$. 
 On the same plot we present curves for pure hadronic  and quark star for reference.  The dash–dot curve represents the hadronic shear viscosity contribution, while the short-dashed  and dash–dash–dot curves correspond to the hadronic bulk viscosity in the asymptotic regimes $f\ll 1$ and $f\gg 1$, respectively. The dot dashed curve represents the quark shear viscosity contribution, whereas the long-dashed and long dash dot curves correspond to the quark bulk viscosity in $f\ll1$, $f\gg1$ regimes, respectively. 

For the two-layer hybrid star model, the contribution from $\eta$ is represented by the solid  blue band, while the short-dashed green band and dotted red band correspond to the  regimes, respectively. The circle represent the observational data points of the millisecond pulsars in the figure \cite{10.1093/mnras/stw3201, 10.1111/j.1365-2966.2012.21171.x}.    The minima of $\Omega/\Omega_K$, calculated using the mean values of the coefficients, are found to be $\left(\Omega/\Omega_K\right)_{\mathrm{min}} = 0.069$ at a temperature $T = 0.256~\mathrm{MeV}$ for the star with mass $1.5\,M_{\odot}$, whereas for the more massive star with $1.75\,M_{\odot}$ the minimum occurs at $\left(\Omega/\Omega_K\right)_{\mathrm{min}} = 0.071$ at $T = 0.234~\mathrm{MeV}$. Consequently, the minima of the instability curves for the two-layered hybrid stars lie between those obtained for a purely hadronic  star and for a pure quark star.  The dynamic instability curves are compared with the timing data of rotating millisecond  pulsars \cite{10.1093/mnras/stw3201, 10.1111/j.1365-2966.2012.21171.x},  with rotational frequencies $f>100$ Hz. We also observe from Fig.~(\ref{rmode}) that the minimum of the instability curve remains unchanged for both the upper and lower bounds of the inferred coefficients. Below, we present the results thus obtained in the tabular form,
\begin{table}[h]
\centering

\begin{tabular}{|l|c|c|c|c|c|c}
\hline
 & \textbf{M ($M_\odot$)} & \textbf{R (km)} & \textbf{T (MeV)} & \boldmath$\Omega / \Omega_k$  \\ 
\hline
Hadronic matter &  1.76 & 12.20 & 5.944 & 0.0347  \\ 
\hline

\multirow{2}{*}{Hybrid}  &  1.5 & 11.00
& $2.589\times10^{-1}$ & 0.0619 \\

 & 1.75 & 10.3& $2.343\times10^{-1}$ & 0.0709  \\ \hline

Quark matter &  1.98 & 11.9 & $1.21\times10^{-4}$ & 0.129 1\\ 
\hline
\end{tabular}

\caption{Summary of intersection values}
\end{table}

Furthermore, it is important to note that  the instability windows obtained for the two-layer hybrid star configurations through the Bayesian inference framework predict the stability  of the LMXB sources XTE J0929$-$314 and XTE J1807$-$294, as well as the millisecond radio pulsars J0437$-$4715 and J2124$-$3358. The locations of these sources relative to the inferred instability curves suggest that the enhanced viscous dissipation arising from the mixed hadron–quark phase provides sufficient damping to stabilize the r-mode oscillations.

\section{Conclusion}\label{sec4}

In this paper we have obtained  semi-analytical expressions for the shear and bulk viscous damping time scales of r-mode oscillations and the corresponding instability boundaries.  We construct a two-layer hybrid-star model and employ Bayesian inference framework by observational LMXB and radio data to predict the dimensionless constants $\tilde{S}$,$\tilde{V}$,$\tilde{W}$ and $\tilde{J}$ which encode both equilibrium and non-equilibrium properties of
the stellar interior. 

The observationally inferred posteriors indicate that the  distributions are relatively compact, indicating that the Bayesian inference framework provides meaningful constraints on the viscous transport constants governing the r-mode dissipation mechanisms. The minima of the instability curve obtained from the inferred posterior further indicates that the temperature at which the minimum occurs remains essentially unchanged when varying the equilibrium and viscous parameters within their inferred mean values and credible intervals.  The morphology of the instability regions suggests that radio pulsar and LMXB observations of XTE J0929–314, XTE J1807–294, J0437–4715, and J2124–3358 are compatible with the two-layer mixed-phase model. These results indicate that a Bayesian approach inferring microscopic predictions of viscous and equilibrium properties, and thus with the determination of the instability minima, provides an effective framework for assessing the stability of hybrid stars.

In future work, we plan to extend this methodology to a more realistic three-layer model in order to improve the physical description of hybrid stars. We also aim to generalize the present formalism to derive quantitative constraints on the  $\eta$ and the $\zeta$, associated with each individual layer of the stellar interior. Furthermore, we aim to characterize the nature of the hadron–quark phase transition—specifically, to determine whether it is of first order, second order, or manifests as a smooth crossover—by performing statistical inference on transport coefficients, constrained by gravitational-wave signatures generated by r-mode oscillations.

\section{Acknowledgments}
 One of the authors, S.S, would like to acknowledge the fruitful discussion with Dr. T Malik regarding the computational framework adopted in this work.


\bibliographystyle{sn-mathphys-num}
\bibliography{sn-bibliography}


\end{document}